\documentclass[runningheads]{llncs}
\usepackage{commath}
\usepackage[table,xcdraw]{xcolor}
\usepackage{enumitem,kantlipsum}
\usepackage{tikz}
\usetikzlibrary{quantikz2}
\usepackage{graphicx}
\usepackage{dcolumn}
\usepackage{bm}
\usepackage{physics}
\usepackage{xcolor}
\usepackage[caption=false]{subfig}
\usepackage{xurl} 
\usepackage[hidelinks]{hyperref}
\usepackage{tabularray}
\usepackage{wrapfig}

\usepackage{tabularx}
\usepackage{array}
\definecolor{workflowgray}{HTML}{C7C7C7}
\definecolor{lightgraycol}{HTML}{F1F1F1}
\definecolor{cobyla}{HTML}{A7BED0}
\definecolor{qng}{HTML}{86A889}
\definecolor{decolor}{HTML}{D1B083}

\begin{document}
\title{
Distributed VQE: Embarrasingly Parallel strategies on NISQ}

\author{M. Losada\inst{1,2} \and D. Fa\'ilde\inst{1} \and A. G\'omez\inst{1}\orcidID{0000-0001-7272-8488}}

\date{\today}
\institute{Galicia Supercomputing Center (CESGA), Avenida de Vigo, s/n, Santiago de Compostela, E-15705, A Coruña, Spain \and
Departamento de Física de Partículas Universidade de Santiago de Compostela, E-15782, A Coruña, Spain}

\titlerunning{Distributed Variational Quantum Eigensolver}
\authorrunning{M. Losada et al.}
\maketitle
\begin{abstract}

Variational Quantum Eigensolver requires many circuit executions, making it ideal for distributed parallelization. However, heterogeneous noise in NISQ devices can skew results and efficiency. Using the CUNQA platform for emulation of virtual QPUs, we evaluate three embarrassingly parallelization strategies (shot-level, circuit-level for gradients and observables and candidate level for population-based optimizers) across metrics like speedup and accuracy. 
\keywords{Distributed VQE \and HPQC \and Noise \and CUNQA}
\end{abstract}


\section{Introduction}

Current quantum devices are still constrained by the limitations of the Noisy
Intermediate-Scale Quantum (NISQ) era: reduced number of qubits, imperfect gates,
restricted connectivity, finite coherence times, and readout errors
\cite{preskill2018nisq}. One route towards scalability is Distributed Quantum Computing
(DQC), where several smaller quantum processing units (QPUs) are considered as part of a
larger computational framework~\cite{DQCReview}. In the current context, where quantum communication
between QPUs is not yet generally available, it is particularly relevant to investigate strategies in which independent quantum tasks are executed separately and their results are combined through classical post-processing.

Variational Quantum Algorithms (VQAs) are well suited to this latter setting.
They rely on a hybrid quantum-classical loop in which a parametrized quantum circuit is
executed on a QPU and a classical optimizer updates its parameters according to a cost
function~\cite{cerezo2021vqa}. The Variational Quantum Eigensolver (VQE) is a
representative example, whose goal is to approximate the ground-state energy of a
Hamiltonian by minimizing the energy expectation value of a parametrized trial state
\cite{peruzzo2014vqe}.

VQE’s high demand for circuit evaluations makes it ideal for parallel execution. However, in NISQ scenarios, hardware noise heterogeneity compromises quantum tasks executions. Since QPUs have varying error profiles, the way tasks are assigned can significantly impact execution time, optimization trajectories, and final accuracy.

The goal of this work is to study embarrassingly parallel strategies for VQE in realistic
distributed NISQ environments with heterogeneous noise profiles using
CUNQA, a DQC emulator designed for HPC environments that supports configurable virtual
QPUs and backend-dependent noisy simulations~\cite{cunqa}. We compare three distributed
workflows: shot-level parallelization, circuit-level parallelization associated with
observables and gradient-related evaluations, and population-level parallelization for a
gradient-free optimizer.

\section{Background}
\label{sec:background}
\subsection{Variational Quantum Eigensolver}

The Variational Quantum   is a hybrid quantum-classical algorithm designed to
estimate the ground-state energy of a physical system. A quantum processor prepares a
parametrized trial state $\ket{\psi(\boldsymbol{\theta})}=U(\boldsymbol{\theta})\ket{\psi_0}$,
where $U(\boldsymbol{\theta})$ is a parametrized quantum circuit and
$\boldsymbol{\theta}$ is the vector of variational parameters. The classical processor
updates these parameters in order to minimize the energy expectation value
\begin{equation}
    E(\boldsymbol{\theta})
    =
    \bra{\psi(\boldsymbol{\theta})}H\ket{\psi(\boldsymbol{\theta})}.
\end{equation}
By the variational principle, this quantity is an upper bound to the exact ground-state
energy. Therefore, the VQE solution is obtained as $E_{\mathrm{VQE}}=\min_{\boldsymbol{\theta}} E(\boldsymbol{\theta})$.

In this work, the target Hamiltonian is the one-dimensional Transverse Field Ising Model
(TFIM)~\cite{Pfeuty1970},
\begin{equation}
    H =
    -J \sum_i^N Z_i Z_{i+1}
    -
    \Gamma \sum_i^N X_i = H_{zz}+H_x,
    \label{eq:tfim}
\end{equation}
with periodic boundary conditions, $J=\Gamma=1$, and $N=6$ spins for this work. The one-dimensional
TFIM is a convenient benchmark because it is analytically solvable for an arbitrary system
size and, for the small system considered here, the exact ground-state can also be
obtained by classical diagonalization. In addition, suitable variational circuits are
known to prepare its ground state accurately for relatively large spin chains, making the
model useful for testing optimizer behaviour and distributed VQE workflows
\cite{av-lm-nat-grad}.

The variational circuit is built using the Hamiltonian Variational Ansatz \cite{HVA} (HVA). For the
TFIM, the interaction ($H_{zz}$) and transverse-field ($H_x$) contributions form two commuting groups,
leading to the layered unitary
\begin{equation}
    U(\boldsymbol{\theta})
    =
    \prod_{\ell=1}^{N_L}
    e^{-iH_{zz}\theta_{2\ell-1}}
    e^{-iH_x\theta_{2\ell}},
    \label{eq:hva}
\end{equation}
where $N_L$ is the number of ansatz layers, which for this work we have chosen to be $N_L=3$\cite{localminima}. Thus, the circuit contains two variational
parameters per layer. The initial state is chosen as
$\ket{\psi_0}=\ket{+}^{\otimes n}$, which is the ground state of the transverse-field
term and can be prepared with a layer of Hadamard gates.

The energy is estimated by measuring the expectation values of the Pauli terms in
Eq.~\eqref{eq:tfim} which correspond to qubit-wise commuting (QWC) groups \cite{tilly}:
elements of $H_{zz}$ can be measured directly in the
computational basis, while elemts from $H_x$ need of Hadamard rotations to measure in the $X$-basis.

\subsection{Optimizers}

The choice of optimizer strongly affects VQE performance, especially under noise \cite{opts-nisq}. In
this work, three representative optimizers are considered. COBYLA \cite{COBYLA} is used as a local
gradient-free baseline. Quantum Natural Gradient \cite{qng}(QNG) uses geometric information from
the quantum state space through the Fubini--Study metric or, equivalently, the real part
of the Quantum Geometric Tensor (QGT). Differential Evolution \cite{DE,de-local-minima}(DE) is a population-based global
optimizer in which several candidate parameter vectors are evaluated at every generation.
Together, these methods allow us to analyze three complementary embarrassingly parallel
strategies.

\subsection{VQAs in noisy environments}

Noise is one of the main limitations for VQE executions on NISQ hardware.
Coherent gate errors, decoherence, dephasing, and readout inaccuracies affect the
prepared state and can significantly degrade the final energy estimate
\cite{georgopoulos2021noise}.

Noise can strongly impair VQA trainability. Barren plateaus, where gradients vanish exponentially with system size, affect both gradient-based and gradient-free optimization \cite{mcclean2018barren,holmes2022expressibility,arrasmith2021gradientfree}. Moreover, noise-induced
barren plateaus (NIBP) have been shown to arise with circuit depth and are described as conceptually
different from the noise-free BP \cite{wang2021noiseinduced}. As a result, mitigation strategies designed to avoid regular BP may not be sufficient in noisy environments.

In distributed VQE executions these effects become more subtle: circuits, shots, or
population candidates may be assigned to QPUs with different noise profiles, combining
estimates affected by different backend-dependent errors. Consequently, the optimizer
does not interact with a single noisy realization of the cost function, but with values
whose noise contribution may depend on the task-to-QPU assignment. Understanding this
interplay between noise, task distribution, and optimizer behaviour is therefore
essential for assessing embarrassingly parallel VQE in realistic NISQ environments,
where the choice of optimizer can strongly affect both convergence and final accuracy
\cite{opts-nisq}.

\subsection{Distributed Quantum Computing}

In Distributed Quantum Computing (DQC), multiple QPUs are used as part of a larger
computational stack, often in an HPC setting where quantum hardware acts as an
accelerator alongside CPUs/GPUs~\cite{QuantumAccelerated,DQCReview}. This integration is
particularly natural for variational algorithms: the quantum device evaluates the cost
function, while classical resources orchestrate the optimization loop and, when several
QPUs are available, can schedule independent quantum tasks in parallel.

Following~\cite{DQCReview}, DQC approaches can be grouped into: (i) \textit{circuit
 distribution}, which executes a single large circuit across devices and requires
quantum and classical communication networks; (ii) \textit{circuit cutting/knitting}
\cite{circuit-cutting}; and (iii) \textit{embarrassingly parallel}
execution, where independent tasks are offloaded to different QPUs and aggregated
classically.

Here we focus on embarrassingly parallel VQE under heterogeneous noise, studying
shot-level, circuit-level, and candidate-level task distribution and their impact on
execution time, parallel efficiency, convergence, and final accuracy.

\section{Parallelization workflows}

This section describes the three embarrassingly parallel workflows considered in this
work. In all cases, the optimizer remains responsible for proposing new
parameters and generating the needed quantum tasks, then those are distributed and
gathered before the next optimization step. The strategies differ in the granularity of
the distributed task: shots, circuits.

\begin{figure}
    \centering
    \subfloat[Shot-level\label{fig:shots-dist}]{%
        \includegraphics[width=0.32\linewidth]{ 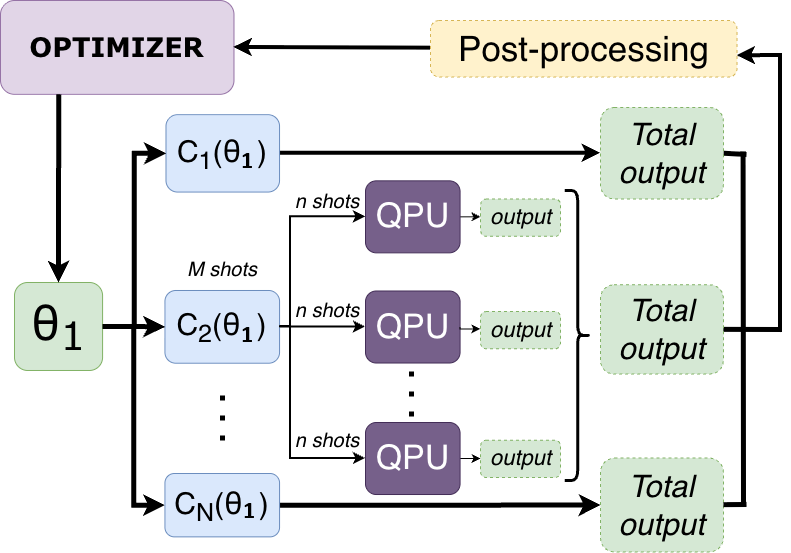}%
    }\hfill
    \subfloat[Circuit-level\label{fig:circuits-wf}]{%
        \includegraphics[width=0.32\linewidth]{ 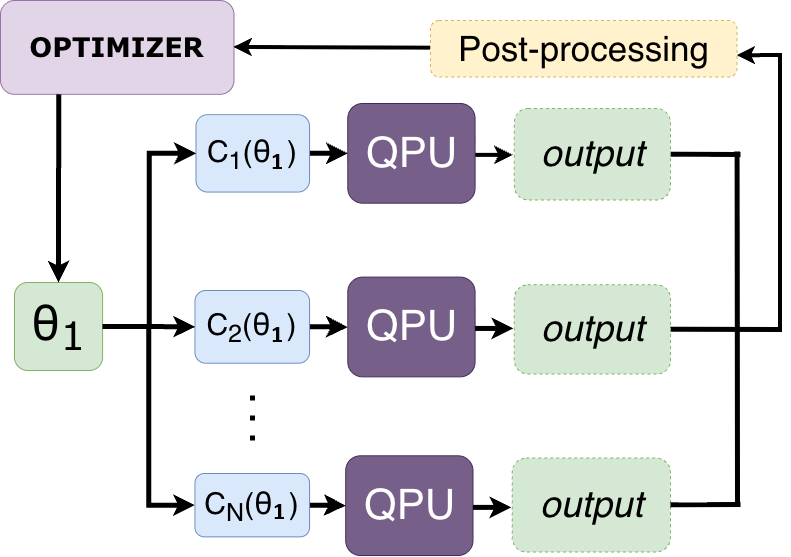}%
    }\hfill
    \subfloat[Parameter-candidate \label{fig:candidates-wf}]{%
        \includegraphics[width=0.32\linewidth]{ 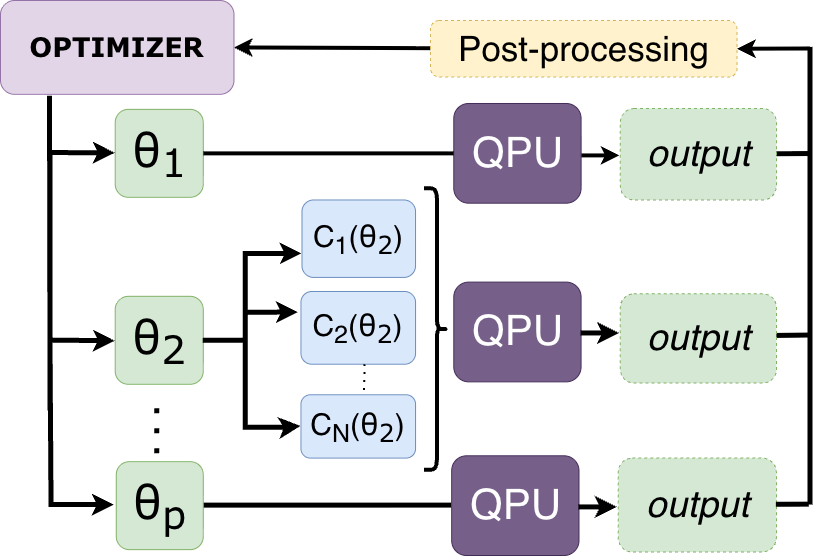}%
    }
    \caption{Workflows for the three embarrassingly parallel distribution strategies.}
    \label{fig:dist-workflows}
\end{figure}

\subsection{Shot-level distribution}

Shot-level distribution can be applied to any circuit evaluation in the VQE loop. For
each circuit $C_i(\boldsymbol{\theta})$, the total shot budget $M$ is split into batches
and executed in parallel across the available QPUs (Fig.~\ref{fig:shots-dist}), so that
each quantum unit runs the same circuit with only a fraction of the samples. The resulting partial counts are put together and then expectation values are computed from the aggregated distribution as in a single-QPU run. This strategy maximizes resources utilization even when few independent circuits are
available, but in heterogeneous pools it mixes samples from different noise profiles,
so the estimator can depend on the shot-to-QPU allocation.

\subsection{Circuit-level distribution}

For a fixed parameter vector $\boldsymbol{\theta}$, the VQE workflow may
require several independent circuit executions before a single optimizer update can be
performed; here the parallel unit is a complete circuit rather than a
subset of shots. As shown in Fig.~\ref{fig:circuits-wf}, these circuits can be assigned to
different QPUs, executed independently, and gathered afterwards in the post-processing
stage. This situation appears naturally in the evaluation of Hamiltonians decomposed into
different terms or measurement groups, for which the energy (cost) associated with a given parameter vector is reconstructed as the sum of the evaluations of many observables that can each land on a different quantum unit. The quantum tasks required depend on how the optimizer carries out the parameter update.

COBYLA proposes a single parameter vector per optimization step, therefore a single cost evaluation. For this work, that makes a total of 2 circuit evaluations per optimization step.

DE produces a configurable-size population of candidate parameter vectors,
each requiring its own cost evaluation. For this case, that is $2\times6(\text{num. param.})\times$ \emph{popsize}(configurable)$=12\times$\emph{popsize} circit evaluations.

QNG, apart from the energy evaluation of the current parameter vector, also requires of gradient-related and metric-related quantities. Gradient estimation in VQAs generally requires multiple circuit evaluations per optimization step, resulting in a substantial computational overhead \cite{GenPSR,Schuld2019}.
 In addition, the number of circuits needed for estimating QGT elements depends strongly on the ansatz
structure. Taking the block-diagonal approximation of the tensor \cite{qng} and HVA, QNG needs a total of $2\times2(\text{energy shifts})\times6(\text{num. param.})\times6(\text{param. multiplicity})+ 6(\text{QGT diagonal}) = 150$ circuit evaluations.

In heterogeneous pools, a fixed circuit-to-QPU mapping may repeatedly associate
specific observables, shifted energies, or QGT terms with the same backend noise profile. For this reason, we consider an ordered permutation of task assignment across
optimization steps.


\subsection{Parameter-candidate distribution}

In this workflow (Fig. \ref{fig:candidates-wf}), designed for population-based optimization methods as Differential Evolution, the parallel unit is a complete
candidate parameter vector. The optimizer generates a population
$\{\boldsymbol{\theta}_1,\ldots,\boldsymbol{\theta}_p\}$ and each candidate is assigned to
a QPU for the evaluation of the cost function. The main difference with the circuit-level strategy applied to DE is that here all
circuits from the same individual are executed on the same QPU.
Thus, each quantum unit is employed for the cost estimate of one candidate.

In heterogeneous configurations, a fixed candidate-to-QPU assignment may systematically
favor or penalize individuals depending on the backend noise profile. For this reason, we
also consider the above mentioned permutation strategy across generations.

\section{Implementation details}
The simulations presented in this work were carried out using CUNQA within the HPC
infrastructures of CESGA. CUNQA provides a framework to emulate distributed quantum
computing workflows through virtual QPUs (vQPUs) deployed as independent computational
resources~\cite{cunqa}. Each vQPU is associated with a specific backend, including gates, connectivity, and a
noise model derived from calibration data. We build heterogeneous pools using backend
data from three IBM (Berlin, Boston, Marrakesh) and one OQC (Qmio~\cite{Qmio}) quantum computers. Calibrations correspond to June 2026. The choice of simulator for the vQPUs was Qiskit AerSimulator, and each of them was deployed on one compute node (64 CPU cores,
15~GB per core), emulating a co-located QPU--HPC integration model~\cite{cunqa}.

A custom Python library, integrated with CUNQA, manages the VQE workflow by asynchronously distributing tasks across vQPUs. It then processes measurement results into energies, gradients, or fitness values before each classical optimizer update.
The COBYLA and Differential Evolution optimizers are implemented via \texttt{scipy.optimize}.
The Quantum Natural Gradient
optimizer is implemented explicitly for this work using Qiskit, CUNQA, and NumPy.

To ensure \emph{reproducibility}, a global seed is propagated across all stages, from transpilation to noisy vQPU executions. This prevents stochastic fluctuations from masking the actual effects of parallelization strategies or noise profiles, allowing for fair comparisons between configurations.

Regarding optimization details, parameters are all initialized around $\pi/5$ \cite{hvanoBP} and the number of shots is set to $10^4$, unless otherwise stated.

\section{Results}
\label{sec:results}
This section compares the three distributed VQE workflows introduced above: shot-level,
circuit-level, and candidate-level distribution. The results are analyzed from two
complementary perspectives. First, we study the impact of the distribution strategy on
the optimization path and final VQE accuracy. Second, we evaluate the speedup and
parallel efficiency.

\subsection{Optimization paths under heterogeneous noise}

\begin{figure}
    \centering
    \includegraphics[width=1\linewidth]{ 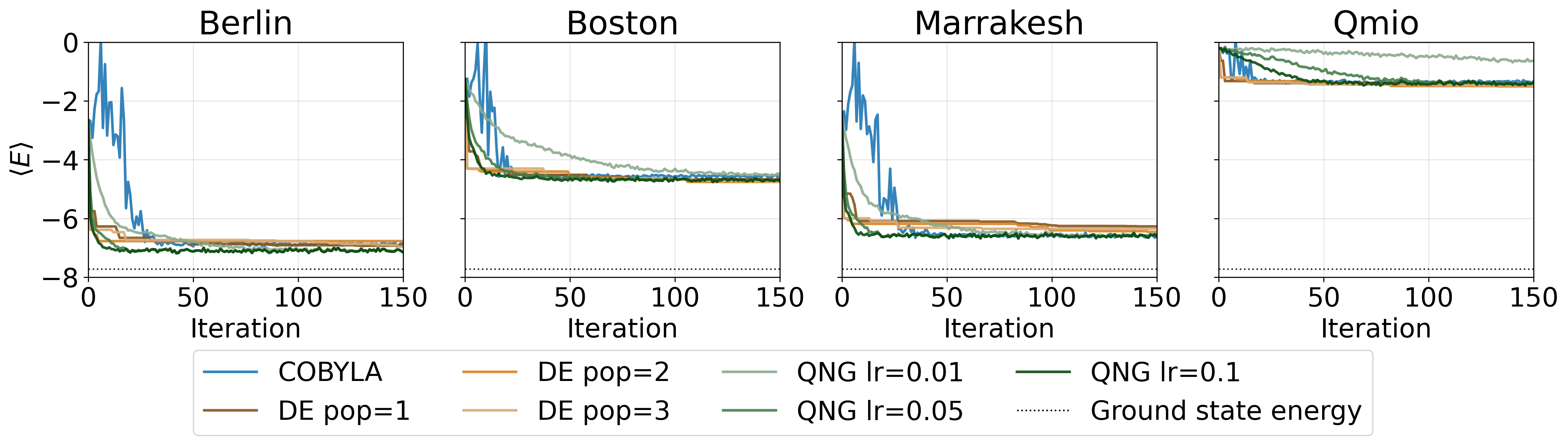}
    \caption{Reference optimization paths for single vQPU executions.}
    \label{fig:optimization-paths-1vqpu}
\end{figure}

Fig.~\ref{fig:optimization-paths-1vqpu} shows single vQPU optimizations using the four mentioned backends.
Due to noise, all optimizers converge above the exact ground-state energy,
shown by the dotted line. As circuit depth increases, noise can flatten the energy
landscape and induce vanishing gradients, limiting further
improvement~\cite{wang2021noiseinduced}. This plots give us an idea of how noisy each device is: IBM Berlin appears to exhibit more favorable convergence behavior, likely due to its longer coherence times and lower gate error rates, whereas Qmio shows less favorable convergence under the opposite hardware conditions. 

Different configurations of \emph{popsize} (pop) and \emph{learning rate} (lr) for DE and QNG were swiped, since these hyperparameters are strongly dependent on the problem it is convenient to do so. Fig. \ref{fig:optimization-paths-1vqpu} shows good behaviour for a \emph{learning rate} of $0.1$ for QNG, whereas the convergence speed of DE does not appear to depend strongly on the population size for this particular case. Given these circumstances, for the parallelization performance study we set a fixed value for these parameters.

\begin{figure}
    \centering
    \includegraphics[width=0.85\linewidth]{ 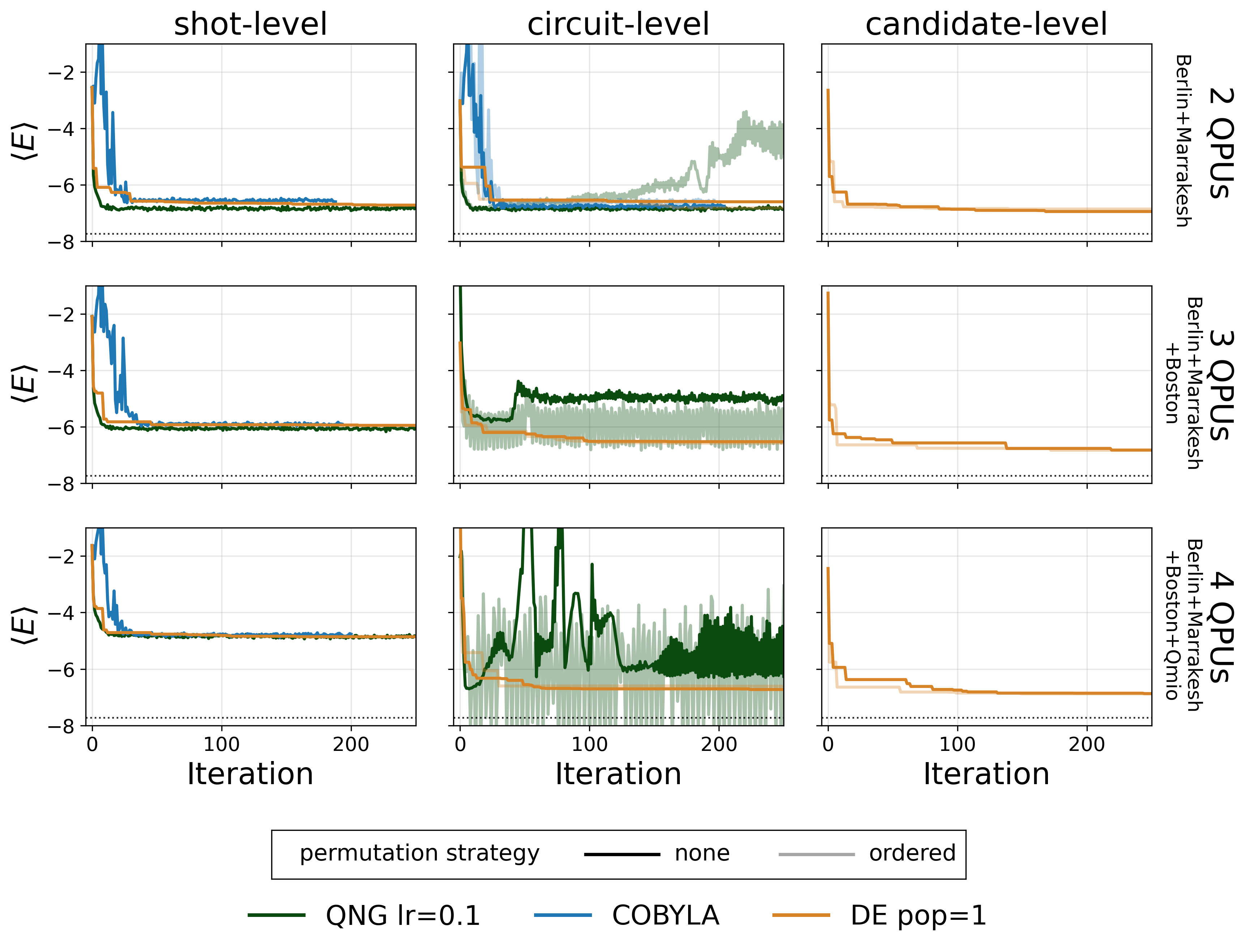}

    \caption{Optimization paths obtained with 2, 3, and 4 vQPUs for the three distributed VQE workflows. The dotted horizontal line marks the reference ground-state energy.}
    \label{fig:optimization-paths-2-4vqpus}
\end{figure}

Fig.~\ref{fig:optimization-paths-2-4vqpus} compares distributed optimizations with progressively more, and increasingly lower-quality, vQPUs. The achieved energy generally worsens as additional backends are included. Shot-level distribution results are actually a clear reflection of this: since counts are aggregated and
the outcome is largely determined by the average device quality, all optimizers are similarly affected.

Circuit-level distribution makes QNG particularly sensitive to lower-quality backends. Since QNG can become unstable when the local geometry is ill-conditioned or the effective step is too large~\cite{yamamoto2019natural}, noise in the comparatively few QGT evaluations may strongly distort the update, leading to unstable or oscillatory convergence, as shown in Fig.~\ref{fig:optimization-paths-2-4vqpus}. Ordered permutation can further aggravate this effect by repeatedly assigning a larger fraction of QGT circuits to the poorest backend.

For COBYLA, circuit-level distribution is only considered for the 2 vQPUs case: it needs of a single cost evaluation, which requires as many circuit evaluations as QWC observables are present in the energy Hamiltonian (\ref{eq:tfim}). Its behaviour is consistent with the rest of the optimizers.

DE remains resilient under circuit-level distribution because its selection mechanism can reject candidates whose cost estimates are degraded by circuits executed on poorer backends. A similar effect occurs for candidate-level distribution, where only a subset of individuals is directly affected, although no substantial improvement in convergence speed is observed. In both cases, permutation redistributes backend-induced bias across the population, preventing the same candidates from being persistently affected by poorer backends. This allows more reliable cost estimates to enter the selection process and leads to a slight overall improvement in convergence speed.

\subsection{Speedup and parallel efficiency}

The speedup and efficiency results indicate that the benefit of distribution depends on
task granularity. For $N$ vQPUs, these are computed as
\begin{equation}
    S_N = \frac{T_1}{T_N},\quad \eta_N = \frac{S_N}{N}
\end{equation}
where $T_1$ is the time per iteration for a single vQPU optimization, while $T_N$
is the one for a $N$-vQPUs optimization. The ideal scaling corresponds to $S_N=N$ and $\eta_N=1$.

The number of shots and candidates is varied to assess the effect of task granularity. No additional levels are considered for circuit distribution, since all circuits constitute equivalent distribution units.

\begin{figure}
    \centering
    \includegraphics[width=\linewidth]{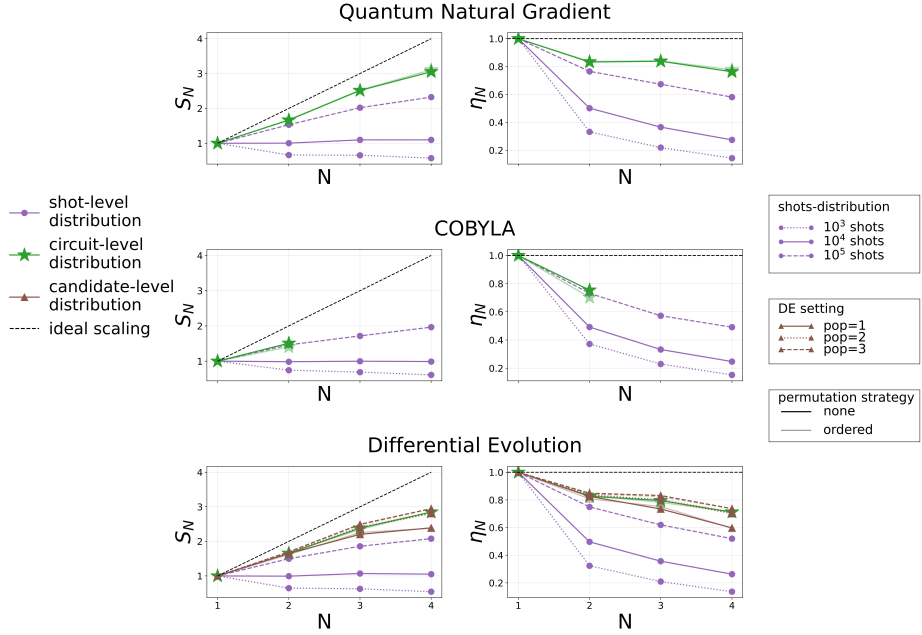}
    \caption{Speedup and parallel efficiency for optimizers under the distributed workflows.}
    \label{fig:speedup-efficiency}
\end{figure}

Fig. \ref{fig:speedup-efficiency} shows that shot-level distribution scales poorly at low sampling budgets across all optimizers, due to distribution and aggregation overhead dominating execution time. Since finite-shot sampling is a major VQA bottleneck \cite{preskill2018nisq}, the improved scaling at higher shot counts suggests that parallelization is most beneficial in measurement-intensive regimes. When overhead remains dominant, measurement-efficient techniques may offer a better alternative (see Section \ref{sec:related-work}).

Circuit-level distribution performs well for both QNG and DE, since each optimization step requires several independent circuit evaluations that can be executed in parallel. QNG shows slightly better scaling, as expected, because it requires a larger number of circuits per step in our setup. However, as discussed in the previous section, this strategy may be particularly sensitive to backend heterogeneity. For COBYLA, the analysis is limited to two vQPUs; within this restricted range, a slight improvement is observed, suggesting that circuit-level distribution could become more beneficial for problems requiring a larger number of circuit evaluations.

Candidate-level distribution becomes more effective as the DE population size increases, since a larger number of independent candidates can be evaluated concurrently, improving vQPU utilization, speedup, and efficiency.

\section{Related work}
\label{sec:related-work}

Several previous works have explored distribution workflows related to those studied here. For instance, at the shot level, reliability-aware frameworks allocate sampling batches according to device calibration \cite{Bisicchia2025ShotWise}, while measurement-efficient methods reduce the required sampling when distribution overhead remains dominant \cite{Kubler2020adaptiveoptimizer,Arrasmith2020operatorsampling}. On the other hand, at the circuit level, QUDIO and Shuffle-QUDIO distribute VQE subproblems across quantum--classical nodes and report runtime acceleration through parallel circuit execution \cite{Du2021QUDIO,ShuffleQUDIO}. More recently, Graph-VQE combines parallel Hamiltonian-term evaluation with block-wise parameter optimization and considers both gradient-free and gradient-based optimizers \cite{Feng2026GraphVQE}. These results are consistent with the scaling observed in this work, showing that circuit-level distribution becomes increasingly effective when multiple independent evaluations are available. Other works have explored more elaborate optimizer-level strategies, such as parameter-parallel training with noise-aware alternation \cite{Niu2023PPDVQA} and multi-population evolutionary schemes in which QPU-specific populations periodically exchange their best solutions \cite{Schiavello2025EvolutionaryQAOA}. These approaches further demonstrate the potential of exploiting optimizer-internal parallelism beyond the direct distribution of candidate evaluations.

In general, several works have proposed flexible frameworks for orchestrating VQA workloads across multiple quantum processors, including quality-aware task assignment on heterogeneous devices \cite{Stein2022EQC} and the distribution of hybrid workloads across local or remote accelerators \cite{Nguyen2024MassivelyParallel}. Within this landscape, the present work contributes a systematic comparison of distribution levels, since the most suitable strategy depends on the optimizer structure, workload, and backend heterogeneity.

\section{Discussion}
\label{sec:discussion}The results show that the usefulness of each distributed VQE workflow depends on the optimizer, task granularity, and backend noise. Shot-, circuit-, and candidate-level distribution were implemented in CUNQA and evaluated on a TFIM instance in terms of execution time, efficiency, convergence, accuracy, and sensitivity to heterogeneous vQPUs.

Shot-level scaling improves with the sampling budget as execution increasingly amortizes submission and aggregation overheads. Circuit-level distribution is effective for QNG and DE when several independent circuits are available, while candidate-level efficiency improves as the DE population size increases.

Backend heterogeneity also influences convergence. Shot-level results reflect the average device quality, QNG is particularly sensitive to noisy circuit evaluations, and DE is more resilient because its selection mechanism can discard degraded candidates. Permutation generally provides a slight improvement by preventing persistent task assignments to poorer backends.

Overall, the most suitable strategy depends on the optimizer structure, workload, and backend quality rather than solely on the number of vQPUs. CUNQA enables these scenarios to be evaluated under controlled conditions~\cite{cunqa}. Future work should validate these trends on real heterogeneous platforms and investigate adaptive scheduling based on workload and backend characteristics.

\begin{credits}
\subsubsection{\ackname}
This work was possible thanks to the QuantumSpain project, funded by the \textit{Ministerio de Transformación Digital y Función Pública del Gobierno de España} through the call for \textit{QUANTUM ENIA – proyecto Quantum Spain}; by the European Union through the NexGenerationEU recovery plan in the framework of the \textit{Agenda España Digital 2026} and by the Galician Regional Government through \textit{Planes Complementarios de I+D+I con las Comunidades Autónomas} in Quantum Communication; and by grant PID2024-159713OB-I00 funded by MICIU/AEI/10.13039/50110 0011033 and by ERDF/EU. Additionally, this research project was made possible through the access granted by the Galicia Supercomputing Center (CESGA) to its infrastructure Qmio quantum computing infrastructure with funding from the European Union, through the Operational Programme Galicia 2014-2020 of ERDF\_REACT EU, as part of the European Union's response to the COVID-19 pandemic. 
\end{credits}

\bibliographystyle{splncs04}
\bibliography{references}

\end{document}